\documentclass[proceedings]{JHEP3}
\usepackage{amsfonts}
\usepackage{amsmath}
\usepackage{epsfig,multicol}

\newbox\mybox

\newcommand\fverb{\setbox\mybox=\hbox\bgroup\verb}
\newcommand\fverbdo{\egroup\medskip\noindent\fbox{\unhbox\mybox}\ }
\newcommand\fverbit{\egroup\item[\fbox{\unhbox\mybox}]}
\conference{Chaos in the thermodynamic Bethe ansatz}
\abstract{We investigate the discretized version of the thermodynamic Bethe ansatz equation
for a variety of 1+1 dimensional quantum field theories. By computing Lyapunov exponents 
we establish that many systems of this type exhibit chaotic behaviour, in the sense that their orbits through 
fixed points are extremely sensitive with regard to the initial conditions.}

\title{Chaos in the thermodynamic Bethe ansatz}
\author{Olalla Castro-Alvaredo$^\circ$ and Andreas Fring$^\bullet$ \\
$^\circ$ Laboratoire de Physique, Ecole Normale Sup\'{e}rieure de Lyon, \\
$\;$ 46 All\'{e}e d'Italie, 69364 Lyon CEDEX, France\\
$\;$ E-mail: \email{ocastroa@ens-lyon.fr}\\
$^\bullet$  Centre for Mathematical Science, City University, \\
$\;$ Northampton Square, London EC1V 0HB, UK\\
$\;$ E-mail: \email{A.Fring@city.ac.uk}}

\input{tcilatex}

\begin{document}

\section{Introduction}

The thermodynamic Bethe ansatz (TBA) equation \cite{ZamoTBA,KM} is an
important tool in the context of 1+1 dimensional integrable quantum field
theories. It serves to extract various types of informations, such as the
Virasoro central charge of the underlying ultraviolet conformal field theory 
\cite{BPZ}, vacuum expectation values \cite{ZamoTBA,CF} etc. As it is a
nonlinear integral equation, it can be solved analytically only in very few
circumstances. In general, one relies on numerical solutions of its
discretised version 
\begin{equation}
\varepsilon _{A}^{n+1}(\theta )=rm_{A}\cosh \theta
-\sum\limits_{B}\int\limits_{-\infty }^{\infty }d\theta ^{\prime }\varphi
_{AB}(\theta -\theta ^{\prime })\ln \left( 1+e^{-\varepsilon _{B}^{n}(\theta
^{\prime })}\right) \,\,.  \label{TBA}
\end{equation}
Here $r$ is the inverse temperature, $m_{A}$ is the mass of a particle of
type $A$, $\theta $ is the rapidity and $\varphi _{AB}$ denotes the
logarithmic derivative of the scattering matrix between the particles of
type $A$ and $B$. The unknown quantities in these equations are the
pseudo-energies $\varepsilon _{A}(\theta )$. The standard solution procedure
for (\ref{TBA}) consists of a consecutive iteration of the equation with
initial values $\varepsilon _{A}^{0}(\theta )=rm_{A}\cosh \theta $. At the
heart of this procedure lie the \textit{assumptions} that the exact solution
is reached for $n\rightarrow \infty $, i.e. the sequence converges, and
furthermore that the final answer is non-sensitive with regard to the
initial values, that is its uniqueness. In general these assumptions are
poorly justified and only few rigorous investigations for some simple models
exist \cite{YY,KM,FKS}. So far the outcome has always been that these
assumptions on the existence and uniqueness of the solution indeed hold,
albeit for certain systems convergence problems in certain regimes have been
noted \cite{CDF,CFInf}. The main purpose of this note is to show that this
believe has to be challenged and is in fact unjustified for certain well
defined theories. We note that our findings do neither effect the principles
of the TBA itself nor the consistency of the quantum field theory it is
meant to investigate. However, they indicate that one needs to be very
cautious when using the above solution procedure and making deductions about
the physics for such theories as one might just be mislead by the
non-convergence of the mathematical procedure used to solve the
TBA-equations.

Here we will not analyze the full TBA-equations (\ref{TBA}), but rather
concentrate on the ultraviolet regime, that is $r\approx 0$, in which it
possess some approximation. Clearly, the occurrence of chaotic behaviour in
this regime will have consequences for the finite temperature regime. We
encounter the interesting phenomenon that the iterative procedure is
convergent beyond the ultraviolet (or for a certain choice of parameters in
some theories), but that unstable fixed points are present in the
ultraviolet, meaning that this regime can never be reached by the iterative
solution procedure for (\ref{TBA}).

\section{Stability of fixed points and Lyapunov exponents}

For completeness, let us first briefly recall a few well known basic facts
concerning the nature of fixed points which may be found in standard text
books on dynamical systems, see e.g. \cite{Dyn}. The objects of our
investigations are difference equations of the type 
\begin{equation}
\vec{x}_{n+1}=\vec{F}(\vec{x}_{n}),  \label{1}
\end{equation}
where $n\in \mathbb{N}_{0}$, $\vec{x}_{n}\in \mathbb{R}^{\ell }$ and $\vec{F}%
:\mathbb{R}^{\ell }\rightarrow \mathbb{R}^{\ell }$ is a vector function. We
are especially interested in the fixed points $\vec{x}_{f}$ of this system
being defined as 
\begin{equation}
\vec{F}(\vec{x}_{f})=\vec{x}_{f}~.
\end{equation}
The fixed point is reached by iterating (\ref{1}), if for a perturbation of
it, defined as $\vec{y}_{n}=\vec{x}_{n}-\vec{x}_{f}$, we have $%
\lim_{n\rightarrow \infty }\vec{y}_{n}=0$. From (\ref{1}) we find 
\begin{equation}
\vec{y}_{n+1}+\vec{x}_{f}=\vec{F}(\vec{y}_{n}+\vec{x}_{f})=\vec{F}(\vec{x}%
_{f})+J\cdot \vec{y}_{n}+\mathcal{O}(\left\vert \vec{y}_{n}\right\vert
^{2})\quad \text{for }\vec{y}_{n}\rightarrow 0  \label{y}
\end{equation}
where $J$ is the $\ell \times \ell $ Jacobian matrix of the vector function $%
\vec{F}(\vec{x})$ 
\begin{equation}
J_{ij}=\left. \frac{\partial F_{i}}{\partial x_{j}}\right\vert _{\vec{x}%
_{f}}~\qquad \qquad \text{for }1\leq i,j\leq \ell .
\end{equation}
The linearized system which arises from (\ref{y}) for $\vec{y}%
_{n}\rightarrow 0$ 
\begin{equation}
\vec{y}_{n+1}=J\cdot \vec{y}_{n}  \label{lin}
\end{equation}
governs the nature of the fixed point under certain conditions \cite{Dyn}.
Evidently, it is solved by 
\begin{equation}
\vec{y}_{n}=q_{i}^{n}\vec{v}_{i}\qquad \quad \text{with \quad }J\cdot \vec{v}%
_{i}=q_{i}^{n}\vec{v}_{i}\quad \text{for }1\leq i\leq \ell ~.  \label{ei}
\end{equation}
Excluding the case when the eigenvectors $\vec{v}_{i}$ of the Jacobian
matrix are not linearly independent, we can expand the initial value
uniquely 
\begin{equation}
\vec{y}_{0}=\sum\limits_{i=1}^{\ell }\varsigma _{i}\vec{v}_{i}
\end{equation}
such that 
\begin{equation}
\vec{y}_{n}=\sum\limits_{i=1}^{\ell }\varsigma _{i}q_{i}^{n}\vec{v}_{i}~.
\label{ex}
\end{equation}
It is now obvious from (\ref{ex}) that the perturbation of the fixed point $%
\vec{y}_{n}$ will grow for increasing $n$ if $\left\vert q_{i}\right\vert >1$
for some $i\in \{1,\ldots ,\ell \}$. In that case the fixed point $\vec{x}%
_{f}$ is said to be linearly unstable. On the other hand, if $\left\vert
q_{i}\right\vert <1$ for all $i\in \{1,\ldots ,\ell \}$ the perturbation
will tend to zero for increasing $n$ and the fixed point $\vec{x}_{f}$ is
said to be linearly stable. It can be shown that under some conditions \cite%
{Dyn} the fixed points are nonlinearly stable when they are linearly stable.

In general, that is for any point $\vec{x}$ rather than just the fixed
points $\vec{x}_{f}$, stability properties are easily encoded in the
Lyapunov exponents $\lambda _{i}$. Roughly speaking the Lyapunov exponents
are a measure for the exponential separation of neighbouring orbits. One
speaks of unstable (chaotic) orbits if $\lambda _{i}>0$ for some $i\in
\{1,\ldots ,\ell \}$ and stable orbits if $\lambda _{i}<0$ for all $i\in
\{1,\ldots ,\ell \}$. For an arbitrary point $\vec{x}$ the $\ell $ Lyapunov
exponents for the above mentioned system (\ref{1}) are defined as 
\begin{equation}
\lambda _{i}=\lim_{n\rightarrow \infty }\left[ \frac{1}{n}\ln \left\vert
q_{i}[\vec{F}^{n}(\vec{x})]\right\vert \right] =\lim_{n\rightarrow \infty }%
\left[ \frac{1}{n}\sum_{k=0}^{n-1}\ln \left\vert q_{i}[\vec{F}^{k}(\vec{x}%
)]\right\vert \right] ,  \label{Lia}
\end{equation}
where the $q_{i}\left( \vec{x}\right) $ are the eigenvalues of the Jacobian
matrix as defined in (\ref{ei}), but now at some arbitrary point $\vec{x}$.
Taking the point to be a fixed point, we can relate (\ref{Lia}) to the above
statements. At the fixed point we have of course $\vec{F}^{k}(\vec{x}_{f})=%
\vec{x}_{f}$, such that 
\begin{equation}
\lambda _{i}=\lim_{n\rightarrow \infty }\left[ \frac{1}{n}%
\sum_{k=0}^{n-1}\ln \left\vert q_{i}\left( \vec{x}_{f}\right) \right\vert %
\right] =\ln \left\vert q_{i}\right\vert ~.
\end{equation}
Therefore, a stable fixed point is characterized by $\lambda _{i}<0$ or $%
\left\vert q_{i}\right\vert <1$ for all $i\in \{1,\ldots ,\ell \}$ and an
unstable fixed point by $\lambda _{i}>0$ or $\left\vert q_{i}\right\vert >1$
for some $i\in \{1,\ldots ,\ell \}$. We can now employ this criterion for
some concrete systems.

\section{Unstable fixed points in constant TBA equations}

We adopt here the notation of \cite{FK,CK,CFq}, by which a large class of
integrable quantum field theories can be referred to in a general Lie
algebraic form as \textbf{g\TEXTsymbol{\vert}\~{g}}-theories. Their
underlying ultraviolet conformal field theories can be described by the
theories investigated in \cite{gg1,gg2,gg3} (and special cases thereof) with
Virasoro central charge $c=\ell \tilde{\ell}\tilde{h}/(h+\tilde{h})$. Here $%
\ell (\tilde{\ell})$ and $h(\tilde{h})$ are the rank and the Coxeter number
of \textbf{g}(\textbf{\~{g}}), respectively. In particular, \textbf{g%
\TEXTsymbol{\vert}A}$_{1}$ is identical to the minimal affine Toda theories
(ATFT) \cite{Toda,DIO} and \textbf{A}$_{n}$\textbf{\TEXTsymbol{\vert}\~{g} }
corresponds to the \textbf{\~{g}}$_{n+1}$-homogeneous Sine-Gordon (HSG)
models \cite{Park:1994bx,Fernandez-Pousa:1997hi}. In this formulation each
particle is labelled by two quantum numbers $(a,i)$, which take their values
in $1\leq a\leq \ell $ and $1\leq i\leq \tilde{\ell}$. Hence, in total we
have $\tilde{\ell}\times \ell $ different particle types. It is a standard
procedure in this context \cite{ZamoTBA,KM} to approximate the
pseudo-energies in (\ref{TBA}) by $\varepsilon _{a}^{i}(\theta )=\varepsilon
_{a}^{i}=$ const in a large region for $\theta $ when $r$ is small. For
convenience one then introduces further the quantity $x_{a}^{i}=\exp
(-\varepsilon _{a}^{i})$ such that (\ref{TBA}) can be cast into the compact
form 
\begin{equation}
x_{a}^{i}=\prod\limits_{b=1}^{\ell }\prod\limits_{j=1}^{\tilde{\ell}%
}(1+x_{b}^{j})^{N_{ab}^{ij}}=:F_{a}^{i}(\vec{x})\qquad \text{with }%
N_{ab}^{ij}=\,\delta _{ab}\delta _{ij}-K_{ab}^{-1}\tilde{K}_{ij}\,.
\label{CT}
\end{equation}
The matrix $N_{ab}^{ij}$ in (\ref{CT}) encodes the information on the
asymptotic behaviour of the scattering matrix. As stated in (\ref{CT}) it is
specific to each of the \textbf{g\TEXTsymbol{\vert}\~{g}}-theories with $K$
and $\tilde{K}$ being the Cartan matrix of \textbf{g} and \textbf{\~{g}},
respectively. The equations (\ref{CT}) are referred to as the constant
TBA-equations. They govern the ultraviolet behaviour of the system and their
solutions yield directly the effective central charge 
\begin{equation}
c_{\text{eff}}=\frac{6}{\pi ^{2}}\sum\limits_{a=1}^{\ell }\sum\limits_{i=1}^{%
\tilde{\ell}}\mathcal{L}\left( \frac{x_{a}^{i}}{1+x_{a}^{i}}\right)
\label{ceff}
\end{equation}
with $\mathcal{L}(x)=\sum_{n=1}^{\infty }x^{n}/n^{2}+\ln x\ln (1-x)/2$
denoting Rogers dilogarithm (see e.g. \cite{Dilog} for properties).

Let us now discretise (\ref{CT}) and analyze it with regard to the nature of
its fixed points. According to the argument of section 2, we have to compute
first of all the Jacobian matrices of $\vec{F}$ 
\begin{equation}
J_{ab}^{ij}=\left. \frac{\partial F_{a}^{i}}{\partial x_{b}^{j}}\right\vert
_{\vec{x}_{f}}=N_{ab}^{ij}\frac{\left( x_{a}^{i}\right) _{f}}{%
1+(x_{b}^{j})_{f}}~.  \label{J}
\end{equation}
Next we need to determine the eigensystem of the Jacobian matrix. This is
not possible to do in a completely generic way at present, since not even
the solutions, i.e. fixed points, of (\ref{CT}) are known in a general
fashion. Instead, we present some examples to exhibit the possible types of
behaviour.

\subsection{Stable fixed points}

We start with a simple example of a stable fixed point. We present the $%
A_{2}|A_{1}$ case, which after the free Fermion ($A_{1}|A_{1}$) is the next
non-trivial example in the series of the minimal ATFTs, the scaling
three-state Potts model with Virasoro central charge $c=4/5$. The TBA has
been investigated in \cite{ZamoTBA,KM}. The constant TBA-equations (\ref{CT}%
) 
\begin{equation}
x_{1}=(1+x_{1})^{-1/3}(1+x_{2})^{-2/3}=F_{1}(\vec{x}),\qquad
x_{2}=(1+x_{1})^{-2/3}(1+x_{2})^{-1/3}=F_{2}(\vec{x}),
\end{equation}
can be solved analytically by the golden ratio $\tau :=(\sqrt{5}-1)/2=$ $%
x_{1}=x_{2}$. Using this solution for the fixed point $\vec{x}%
_{f}=(x_{1},x_{2})$, we compute the Jacobian matrix of $\vec{F}(\vec{x})$ 
\begin{equation}
J(\vec{x}_{f})=-\frac{1}{3}\left( 
\begin{array}{rr}
\tau ^{2} & 2\tau ^{2} \\ 
2\tau ^{2} & \tau ^{2}%
\end{array}
\right) ,
\end{equation}
with eigensystem 
\begin{equation*}
q_{1}=-\tau ^{2}\approx -0.38197,\quad \vec{v}_{1}=(1,1),\quad q_{2}=\tau
^{2}/3\approx 0.12732,\quad \vec{v}_{2}=(-1,1)~.
\end{equation*}
As the eigenvectors are obviously linearly independent and $\left\vert
q_{1}\right\vert <1,\left\vert q_{2}\right\vert <1$, we deduce that all
Lyapunov exponents are negative and therefore that the fixed point is
stable. Indeed, numerical studies of this system exhibit a fast convergence
and a non-sensitive behaviour with regard to the initial values of the
iterative procedure.

For some theories the solutions are known analytically in a closed form. For
instance, the constant TBA equations for the $A_{1}|A_{\ell }$-theories ($%
\equiv SU(\ell +1)_{2}$-HSG-model) are solved by 
\begin{equation}
x_{1}^{i}=\left[ \frac{\sin [\pi (1+i)\lambda ]}{\sin (\pi \lambda )}\right]
^{2}-1,\quad \quad \text{for }1\leq i\leq \ell
\end{equation}
\noindent with $\lambda =1/(3+\ell )$ \cite{Kun1,Kun2,Kun3,CFq}. Taking this
solution for the fixed point we compute the Jacobian matrix (\ref{J}) with $%
N_{ij}=(\delta _{i,j+1}+\delta _{i,j-1})/2$ to 
\begin{equation}
J_{11}^{ij}(\vec{x}_{f})=\frac{1}{2}\left[ \frac{\sin [\pi (2+i)\lambda ]}{%
\sin (i\pi \lambda )}\delta _{i,j+1}+\frac{\sin (i\pi \lambda )}{\sin [\pi
(2+i)\lambda ]}\delta _{i,j-1}\right] .  \label{Ja}
\end{equation}
and the eigenvalues to $q_{i}=\cos [\pi (i+1)\lambda ]$. As $\left\vert
q_{i}\right\vert <1$ for all $i\in \{1,\ldots ,\ell \}$ the fixed points are
stable. We investigated various minimal affine Toda field theories which all
posses fixed points of this nature. In fact, the general assumption is that
all systems exhibit such a behaviour. We present now some counter examples
which refute this believe.

\subsection{Stable two-cycles}

We start with a system which does not possess a stable fixed point, but
rather a stable two-cycle, i.e. a solution for$\ $%
\begin{equation}
\vec{G}(\vec{x}):=\vec{F}(\vec{F}(\vec{x}))=\vec{x}~.  \label{G}
\end{equation}
We consider the $A_{2}|A_{2}$-theories ($\equiv SU(3)_{3}$-HSG model)
studied already previously by means of the TBA in \cite%
{Castro-Alvaredo:1999em}. Its extreme ultraviolet Virasoro central charge
is\ $c=2$. The constant TBA-equations for this case read 
\begin{eqnarray}
x_{1}^{1} &=&\frac{(1+x_{1}^{2})^{2/3}(1+x_{2}^{2})^{1/3}}{%
(1+x_{1}^{1})^{1/3}(1+x_{2}^{1})^{2/3}}=F_{1}(\vec{x}),\quad x_{2}^{1}=\frac{%
(1+x_{1}^{2})^{1/3}(1+x_{2}^{2})^{2/3}}{%
(1+x_{1}^{1})^{2/3}(1+x_{2}^{1})^{1/3}}=F_{2}(\vec{x}),~~  \label{e1} \\
x_{1}^{2} &=&\frac{(1+x_{1}^{1})^{2/3}(1+x_{2}^{1})^{1/3}}{%
(1+x_{1}^{2})^{1/3}(1+x_{2}^{2})^{2/3}}=F_{3}(\vec{x}),\quad x_{2}^{2}=\frac{%
(1+x_{1}^{1})^{1/3}(1+x_{2}^{1})^{2/3}}{%
(1+x_{1}^{2})^{2/3}(1+x_{2}^{2})^{1/3}}=F_{4}(\vec{x}),~~  \label{e2}
\end{eqnarray}
with analytic solution $x_{1}^{1}=x_{2}^{1}=x_{1}^{2}=x_{2}^{2}=1$. Taking
this solution as the fixed point, we compute the Jacobian matrix for $\vec{F}%
(\vec{x})$ 
\begin{equation}
J(\vec{x}_{f})=\frac{1}{6}\left( 
\begin{array}{rrrr}
-1 & -2 & 2 & 1 \\ 
-2 & -1 & 1 & 2 \\ 
2 & 1 & -1 & -2 \\ 
1 & 2 & -2 & -1%
\end{array}
\right)
\end{equation}
with eigensystem 
\begin{equation}
\begin{array}{llll}
q_{1}=-1,\qquad & \vec{v}_{1}=(-1,-1,1,1),\qquad & q_{2}=1/3,\qquad & \vec{v}%
_{2}=(-1,1,-1,1),\qquad \\ 
q_{3}=0, & \vec{v}_{3}=(1,0,0,1), & q_{4}=0, & \vec{v}_{4}=(0,1,1,0).%
\end{array}%
\end{equation}
We observe that there is one eigenvalue with $|q_{1}|=1$, which is generally
called a marginal behaviour, i.e. the stability properties depend on the
other eigenvalues and on the next leading order. In fact, we can see from (%
\ref{ex}) that the perturbation of the fixed point will remain the same even
for large values of $n$, flipping between two values and thus suggesting the
existence of a stable two cycle (\ref{G}). We find that (\ref{G}) can be
solved by 
\begin{equation}
x_{1}^{1}=x_{2}^{1}=1/x_{1}^{2}=1/x_{2}^{2}=\kappa ,  \label{ss}
\end{equation}
for any arbitrary value of $\kappa $. To determine the stability of the
two-cycle we have to compute the Jacobian matrix for $G(\vec{x})$%
\begin{equation}
J(\vec{x}_{f})=\frac{1}{9+9\kappa }\left( 
\begin{array}{rrrr}
5\kappa & 4\kappa & -4\kappa ^{2} & -5\kappa ^{2} \\ 
4\kappa & 5\kappa & -5\kappa ^{2} & -4\kappa ^{2} \\ 
-4/\kappa & -5/\kappa & 5 & 4 \\ 
-5/\kappa & -4/\kappa & 4 & 5%
\end{array}
\right)
\end{equation}
which has eigensystem 
\begin{equation}
\begin{array}{llll}
q_{1}=1,\qquad & \vec{v}_{1}=(-\kappa ^{2},-\kappa ^{2},1,1)\qquad & 
q_{2}=1/9,\qquad & \vec{v}_{2}=(-\kappa ^{2},\kappa ^{2},-1,1),\qquad \\ 
q_{3}=0, & \vec{v}_{3}=(\kappa ,0,0,1), & q_{4}=0, & \vec{v}_{4}=(0,\kappa
,1,0).%
\end{array}%
\end{equation}
We conclude from this that one approaches a stable two-cycle when iterating
the discretised version of (\ref{G}). Thus, we note that the TBA-system for
the $SU(3)_{3}$-HSG model in the ultraviolet regime does not posses a stable
fixed point but an infinite number of stable two-cycles of the type (\ref{ss}%
). It is now intriguing to note that when using this solution to compute the
effective Virasoro central charge (\ref{ceff}), one always obtains the
expected value $c=2$ for any value of $\kappa \in \mathbb{R}$, simply due to
an identity for the Rogers dilogarithm $\mathcal{L}(1-x)+\mathcal{L}(x)=\pi
^{2}/6$. Hence, despite the fact, that one is using entirely wrong
pseudo-energies, one obtains by pure luck an apparent confirmation of the
theories consistency.

\subsection{Unstable fixed points, chaotic behaviour}

In this section we present some TBA-systems for well-defined quantum field
theories, which exhibit a chaotic behaviour in the sense that their
iterative solutions are extremely sensitive with regard to the initial
values.

\subsubsection{$A_{4}|A_{4}$}

This model is the $SU(5)_{5}$-HSG model with extreme ultraviolet Virasoro
central charge\ $c=8$. To reduce the complexity of the model, we exploit
already from the very beginning the $\mathbb{Z}_{2}$-symmetries in the $%
A_{4} $-Dynkin diagrams and identify $%
x_{1}^{1}=x_{1}^{4}=x_{4}^{1}=x_{4}^{4} $, $%
x_{2}^{2}=x_{2}^{3}=x_{3}^{2}=x_{3}^{3}$, $%
x_{2}^{1}=x_{3}^{1}=x_{2}^{4}=x_{3}^{4}$, $%
x_{1}^{2}=x_{4}^{2}=x_{1}^{3}=x_{4}^{3}$. With these identifications the
constant TBA-equations can be brought into the form 
\begin{eqnarray}
x_{1}^{1} &=&\frac{(1+x_{1}^{2})(1+x_{2}^{2})}{(1+x_{1}^{1})(1+x_{2}^{1})^{2}%
}=F_{1}(\vec{x}),\qquad \quad x_{2}^{1}=\frac{(1+x_{1}^{2})(1+x_{2}^{2})^{2}%
}{(1+x_{1}^{1})^{2}(1+x_{2}^{1})^{3}}=F_{2}(\vec{x}),  \label{a1} \\
x_{1}^{2} &=&\frac{(1+x_{1}^{1})(1+x_{2}^{1})}{(1+x_{2}^{2})}=F_{3}(\vec{x}%
),\qquad \quad x_{2}^{2}=\frac{(1+x_{1}^{1})^{2}(1+x_{2}^{1})^{2}}{%
(1+x_{1}^{2})(1+x_{2}^{2})}=F_{4}(\vec{x}).  \label{a2}
\end{eqnarray}
We can solve these equations analytically by one, $\tau $ and $\tilde{\tau}%
:=1/\tau =(\sqrt{5}+1)/2$ 
\begin{eqnarray}
x_{1}^{1}
&=&x_{1}^{4}=x_{4}^{1}=x_{4}^{4}=x_{2}^{2}=x_{2}^{3}=x_{3}^{2}=x_{3}^{3}=1, 
\notag \\
x_{1}^{2} &=&x_{4}^{2}=x_{1}^{3}=x_{4}^{3}=\tilde{\tau},  \label{s1} \\
x_{2}^{1} &=&x_{3}^{1}=x_{2}^{4}=x_{3}^{4}=\tau .  \notag
\end{eqnarray}
With this solution for $\vec{x}_{f}$ at hand we compute the Jacobian matrix
of $\vec{F}(\vec{x})$ in (\ref{a1}), (\ref{a2}) 
\begin{equation}
J(\vec{x}_{f})=\left( 
\begin{array}{rrrr}
-1/2 & -2\tau & \tau ^{2} & 1/2 \\ 
-\tau & -3\tau ^{2} & 2\tau -1 & \tau \\ 
\tilde{\tau}/2 & 1 & 0 & -\tilde{\tau}/2 \\ 
1/2 & 2\tau & -\tau ^{2} & -1/2%
\end{array}
\right) .
\end{equation}
Now we find the eigensystem 
\begin{equation}
\begin{array}{ll}
q_{1}=-\tilde{\tau}^{2}\approx -2.6180,\qquad & \vec{v}_{1}=(-1,-1,1,1), \\ 
q_{2}=4\tau -2\approx 0.47214,\qquad \quad & \vec{v}_{2}=(-1,\tau ^{2},-%
\tilde{\tau}^{2},1), \\ 
q_{3}=0, & \vec{v}_{3}=(1,0,0,1), \\ 
q_{4}=0, & \vec{v}_{4}=(-2\tau ^{2},\tau ,1,0).%
\end{array}
~
\end{equation}
As $|q_{1}|>1$ and the eigenvectors are linearly independent, we deduce that
the Lyapunov exponent $\lambda _{1}$ is positive and therefore that the
fixed point (\ref{s1}) is unstable. Indeed, numerical studies of this system
exhibit that any small perturbation away from the solution (\ref{s1}) will
lead to a divergent iterative procedure.

Nonetheless, by some manipulations of (\ref{a1}), (\ref{a2}) one can find
equivalent sets of equations which posses stable fixed points and can be
solved by means of an iterative procedure. For example, when simply
substituting $x_{1}^{1}$ in $F_{2}(\vec{x})$ we obtain the equations 
\begin{eqnarray}
x_{1}^{1} &=&F_{1}(\vec{x})=F_{1}^{\prime }(\vec{x}),\quad \quad x_{2}^{1}=%
\frac{x_{1}^{1}(1+x_{2}^{2})}{(1+x_{1}^{1})(1+x_{2}^{1})}=F_{2}^{\prime }(%
\vec{x}),\quad  \label{b2} \\
x_{1}^{2} &=&F_{3}(\vec{x})=F_{3}^{\prime }(\vec{x}),\quad \quad
x_{2}^{2}=F_{4}(\vec{x})=F_{4}^{\prime }(\vec{x}),
\end{eqnarray}
which are of this kind. Now all Lyapunov exponents resulting from the
Jacobian matrix for $\vec{F}^{\prime }(\vec{x})$ at the fixed point (\ref{s1}%
) are negative. One should note, however, that even though $\vec{F}^{\prime
}(\vec{x})$ it is easily constructed by trial and error from $\vec{F}(\vec{x}%
)$ for the constant TBA-equations, the equivalent manipulations on the full
TBA-equations (\ref{TBA}) are quite unnatural, albeit not impossible to
perform once (\ref{CT}) is analyzed.

One of the distinguishing features of the HSG-models is that they contain
unstable particles in their spectrum, whose masses are characterized by some
resonance parameters $\sigma _{ij}$ with $1\leq i,j\leq \tilde{\ell}$. We
can now interpret these parameters as bifurcation parameters as common in
the study of chaotic systems and investigate the nature of the fixed points
when these parameters are varied. In \cite{CDF} a precise decoupling rule
was provided, which describes the behaviour of the theories when some of the 
$\sigma ^{\prime }$s become large and tend to infinity. For $SU(5)_{5}$ we
have for instance the following possibilities 
\begin{equation}
\lim_{\sigma _{12}\rightarrow \infty }SU(5)_{5}=SU(2)_{5}\otimes
SU(4)_{5}\quad \text{or\quad }\lim_{\sigma _{23}\rightarrow \infty
}SU(5)_{5}=SU(3)_{5}\otimes SU(3)_{5}~.
\end{equation}%
For the algebras involved we found that the fixed point of $SU(4)_{5}$ is
unstable, whereas the fixed points of $SU(3)_{5}$ and $SU(2)_{5}$ are
stable. For our $SU(5)_{5}$ example this implies that the fixed point in $%
SU(3)_{5}\otimes SU(3)_{5}$ will be stable, whereas the fixed point in $%
SU(2)_{5}\otimes SU(4)_{5}$ will be unstable. In general, we find that,
while approaching the ultraviolet from the infrared, once the nature of the
fixed point has changed from stable to unstable it remains that way. This
behaviour can be encoded naturally in standard bifurcation diagrams, which
we present elsewhere.

We also found unstable fixed points for other HSG-models related to simply
laced algebras and \textbf{g\TEXTsymbol{\vert}\~{g}}-theories which are
neither HSG nor minimal ATFT. A priori the behaviour is difficult to
predict, e.g. whereas $D_{4}|D_{4}$ (see \cite{FK} for the solution of \ (%
\ref{CT}) ) and $D_{4}|A_{4}$ have unstable fixed points, the fixed point in 
$D_{4}|A_{2}$ is stable.

\subsubsection{$A_{1}|C_{2}$}

This model is the simplest example of an HSG model related to a non-simply
laced algebra, namely the $Sp(4)_{2}$-HSG model with central charge $c=2$.
In the TBA analysis carried out in \cite{CDF} convergence problems in the
ultraviolet regime were already commented upon. \noindent In fact, we find
here that all HSG models which are related to non-simply laced Lie algebras
posses unstable fixed points. The constant TBA-equations for $A_{1}|C_{2}$
read 
\begin{eqnarray}
x_{1}^{1} &=&\sqrt{(1+x_{1}^{2})(1+x_{3}^{2})}(1+x_{2}^{2})=F_{1}(\vec{x}%
),~x_{1}^{2}=\frac{1}{(1+x_{2}^{2})}\sqrt{\frac{(1+x_{1}^{1})}{%
(1+x_{1}^{2})(1+x_{3}^{2})}}=F_{2}(\vec{x}),~~~~~~~ \\
x_{2}^{2} &=&\frac{(1+x_{1}^{1})}{(1+x_{1}^{2})(1+x_{2}^{2})(1+x_{3}^{2})}%
=F_{3}(\vec{x}),~~x_{3}^{2}=\frac{1}{(1+x_{2}^{2})}\sqrt{\frac{(1+x_{1}^{1})%
}{(1+x_{1}^{2})(1+x_{3}^{2})}}=F_{4}(\vec{x}),
\end{eqnarray}
\noindent with solutions 
\begin{equation}
x_{1}^{1}=3,\quad x_{1}^{2}=x_{3}^{2}=2/3\quad \text{and}\quad x_{1}^{2}=4/5.
\label{soll}
\end{equation}
\noindent The corresponding Jacobian matrix for $\vec{F}(\vec{x})$ reads 
\begin{equation}
J(\vec{x}_{f})=\left( 
\begin{array}{rrrr}
0 & 9/10 & 5/3 & 9/10 \\ 
1/12 & -1/5 & -10/27 & -1/5 \\ 
1/5 & -12/25 & -4/9 & -12/25 \\ 
1/12 & -1/5 & -10/27 & -1/5%
\end{array}
\right)
\end{equation}
\noindent with eigensystem 
\begin{equation}
\begin{array}{ll}
q_{1}\approx -1.3647,\qquad & \vec{v}_{1}=(-3.53,1,1.8104,1), \\ 
q_{2}\approx 0.3973, & \vec{v}_{2}=(-80.2656,1,-20.2124,1), \\ 
q_{3}\approx 0.1229, & \vec{v}_{3}=(2.1956,1,-0.9180,1), \\ 
q_{4}=0, & \vec{v}_{4}=(0,-1,0,1).%
\end{array}
~.
\end{equation}
\noindent Since $|q_{1}|>0$ we find a positive Lyapunov exponent and
therefore the fixed point $\vec{x}_{f}$ \ in (\ref{soll}) is unstable.

\noindent We also checked explicitly $%
A_{1}|C_{3},A_{1}|C_{4},A_{1}|G_{2},A_{2}|B_{2}$ and found a similar
behaviour. Based on these examples we conjecture that the constant
TBA-equations (\ref{CT}) related to \textbf{g}-HSG-models with \textbf{g}
non-simply laced have unstable fixed points.

\section{Conclusions}

We showed that the discretised TBA-equations for many well-defined quantum
field theories exhibit chaotic behaviour in the sense that their orbits are
extremely sensitive with regard to the initial conditions. In particular, we
found several examples for HSG-models and \textbf{g\TEXTsymbol{\vert}\~{g}}%
-theories which are neither HSG nor minimal ATFT. Apart from the statements,
that all $A_{1}|A_{\ell }$-theories have stable fixed points and apparently
all HSG-models which are related to non-simply laced models have unstable
fixed points, we did not find yet a general pattern which characterizes such
theories in a more concise way.

Our findings clearly explain the convergence problems reported upon earlier
in \cite{CDF} and we stress here that they do neither effect the consistency
of the quantum field theories nor the validity of the principles underlying
the TBA, but only point out the need to solve these theories by alternative
means. The closest would be to alter the iterative procedure for (\ref{TBA})
as indicated in section 3.3.1 for the constant TBA-equations, by defining
equivalent sets of equations which have stable fixed points. Unfortunately,
we can not settle with these arguments the convergence problems for the
models studied in \cite{CFInf}, as for those the fixed points are situated
at infinity.

Our results clearly indicate that one can only be confident about results
obtained from iterating (\ref{TBA}) if the nature of the fixed points is
clarified.

\medskip

\noindent \textbf{Acknowledgments: }This work is supported in part by the EU
network \textquotedblleft EUCLID, \emph{Integrable models and applications:
from strings to condensed matter}\textquotedblright , HPRN-CT-2002-00325.

\bibliographystyle{phreport}
\bibliography{Ref}

\begin{thebibliography}{10}

\bibitem{ZamoTBA}
A.~B. Zamolodchikov,
\newblock Thermodynamic Bethe ansatz in relativistic models: Scaling 3-state
  Potts and Lee-Yang models,
\newblock Nucl. Phys. {\bf B342}, 695--720 (1990).

\bibitem{KM}
T.~R. Klassen and E.~Melzer,
\newblock The Thermodynamics of purely elastic scattering theories and
  conformal perturbation theory,
\newblock Nucl. Phys. {\bf B350}, 635--689 (1991).

\bibitem{BPZ}
A.~A. Belavin, A.~M. Polyakov, and A.~B. Zamolodchikov,
\newblock Infinite conformal symmetry in two-dimensional quantum field theory,
\newblock Nucl. Phys. {\bf B241}, 333--380 (1984).

\bibitem{CF}
O.~Castro-Alvaredo and A.~Fring,
\newblock On vacuum energies and renormalizability in integrable quantum field
  theories,
\newblock Nucl. Phys. {\bf B687}, 303--322 (2004).

\bibitem{YY}
C.-N. Yang and C.~P. Yang,
\newblock Thermodynamics of one-dimensional system of bosons with repulsive
  delta function interaction,
\newblock J. Math. Phys. {\bf 10}, 1115--1122 (1969).

\bibitem{FKS}
A.~Fring, C.~Korff, and B.~J. Schulz,
\newblock The ultraviolet behaviour of integrable quantum field theories,
  affine Toda field theory,
\newblock Nucl. Phys. {\bf B549}, 579--612 (1999).

\bibitem{CDF}
O.~A. Castro-Alvaredo, J.~Drei{\ss}ig, and A.~Fring,
\newblock Integrable scattering theories with unstable particles,
\newblock The European Physical Journal {\bf C35}, 393--411 (2004).

\bibitem{CFInf}
O.~Castro-Alvaredo and A.~Fring,
\newblock Constructing infinite particle spectra,
\newblock Phys. Rev. {\bf D64}, 0850051--0850057 (2001).

\bibitem{Dyn}
P.~G. Drazin,
\newblock Nonlinear Systems,
\newblock CUP, Cambridge, 1992.

\bibitem{FK}
A.~Fring and C.~Korff,
\newblock Colour valued scattering matrices,
\newblock Phys. Lett. {\bf B477}, 380--386 (2000).

\bibitem{CK}
C.~Korff,
\newblock Colour valued scattering matrices from non simply-laced Lie algebras,
\newblock Phys. Lett. {\bf B501}, 289--296 (2001).

\bibitem{CFq}
O.~A. Castro-Alvaredo and A.~Fring,
\newblock Scaling functions from q-deformed Virasoro characters,
\newblock J. Phys. {\bf A35}, 609--636 (2002).

\bibitem{gg1}
D.~Gepner,
\newblock New conformal field theories associated with Lie algebras and their
  partition functions,
\newblock Nucl. Phys. {\bf B290}, 10 (1987).

\bibitem{gg2}
G.~V. Dunne, I.~Halliday, and P.~Suranyi,
\newblock Bosonization of parafermionic conformal field theories,
\newblock Nucl. Phys. {\bf B325}, 526 (1989).

\bibitem{gg3}
J.~M. Camino, A.~V. Ramallo, and J.~M. Sanchez~de Santos,
\newblock Graded parafermions,
\newblock Nucl. Phys. {\bf B530}, 715--741 (1998).

\bibitem{Toda}
A.~V. Mikhailov, M.~A. Olshanetsky, and A.~M. Perelomov,
\newblock Two dimensional generalized Toda lattice,
\newblock Commun. Math. Phys. {\bf 79}, 473 (1981).

\bibitem{DIO}
D.~I. Olive and N.~Turok,
\newblock Local conserved densities and zero curvature conditions for Toda
  lattice field theories,
\newblock Nucl. Phys. {\bf B257}, 277 (1985).

\bibitem{Park:1994bx}
Q.-H. Park,
\newblock Deformed coset models from gauged WZW actions,
\newblock Phys. Lett. {\bf B328}, 329--336 (1994).

\bibitem{Fernandez-Pousa:1997hi}
C.~R. Fernandez-Pousa, M.~V. Gallas, T.~J. Hollowood, and J.~L. Miramontes,
\newblock The symmetric space and homogeneous sine-Gordon theories,
\newblock Nucl. Phys. {\bf B484}, 609--630 (1997).

\bibitem{Dilog}
A.~N. Kirillov,
\newblock Dilogarithm identities,
\newblock Prog. Theor. Phys. Suppl. {\bf 118}, 61--142 (1995).

\bibitem{Kun1}
A.~Kuniba,
\newblock Thermodynamics of the $U_q(X_r(1))$ Bethe ansatz system with q a root
  of unity,
\newblock Nucl. Phys. {\bf B389}, 209--246 (1993).

\bibitem{Kun2}
A.~Kuniba and T.~Nakanishi,
\newblock Spectra in conformal field theories from the Rogers dilogarithm,
\newblock Mod. Phys. Lett. {\bf A7}, 3487--3494 (1992).

\bibitem{Kun3}
A.~Kuniba, T.~Nakanishi, and J.~Suzuki,
\newblock Characters in conformal field theories from thermodynamic Bethe
  ansatz,
\newblock Mod. Phys. Lett. {\bf A8}, 1649--1660 (1993).

\bibitem{Castro-Alvaredo:1999em}
O.~A. Castro-Alvaredo, A.~Fring, C.~Korff, and J.~L. Miramontes,
\newblock Thermodynamic Bethe ansatz of the homogeneous sine-Gordon models,
\newblock Nucl. Phys. {\bf B575}, 535--560 (2000).

\end{thebibliography}

\end{document}